\documentclass{article}

\usepackage{arxiv}

\usepackage{subcaption}
\usepackage{amsmath}
\usepackage{textgreek}
\usepackage{textcomp}
\usepackage{cleveref}
\usepackage{float}
\usepackage{units}

\usepackage{authblk}

\usepackage[utf8]{inputenc} 
\usepackage[T1]{fontenc}    
\usepackage{url}            
\usepackage{booktabs}       
\usepackage{amsfonts}       
\usepackage{nicefrac}       
\usepackage{microtype}      
\usepackage{lipsum}
\usepackage{graphicx}

\graphicspath{ {./images/} }

\title{Semiconductor waveguide quantum well lasers}

\author[1]{Jonathan R. C. Woods}
\author[1]{Jon Gorecki}
\author[2]{Roman Bek}
\author[1]{Stephen C. Richardson}
\author[1]{Jake Daykin}
\author[1]{Grace Hooper}
\author[1]{Emelia Branagan-Harris}
\author[1]{Anne C. Tropper}
\author[3]{James S. Wilkinson}
\author[4]{Michael Jetter}
\author[1]{Vasilis Apostolopoulos}
\affil[1]{School of Physics and Astronomy, University of Southampton, Southampton, SO17 1BJ}
\affil[2]{Twenty-One Semiconductors GmbH, Kiefernweg 4, 72654 Neckartenzlingen, Germany}
\affil[3]{Zepler Institute, University of Southampton, Southampton, SO17 1BJ}
\affil[4]{ Institute for Semiconductor Optics and Functional Interfaces, University of Stuttgart, 70569 Stuttgart, Germany}
\affil[ ]{E-mail: va2w07@soton.ac.uk}

\begin{document}

\maketitle


\begin{abstract}
Coherent laser arrays compatible with silicon photonics are demonstrated in a waveguide geometry in epitaxially grown semiconductor membrane quantum well lasers transferred on substrates of silicon carbide and oxidised silicon; we record lasing thresholds as low as 60\,mW of pump power. We study the emission of single lasers and arrays of lasers in the sub-mm range. We are able to create waveguide laser arrays with modal widths of approximately 5-10 \textmu m separated by 10-20 \textmu m, using real and reciprocal space imaging we study their emission characteristics and find that they maintain their mutual coherence while operating on either single or multiple longitudinal modes per lasing cavity. 
\end{abstract}

\section{Introduction}

During at least the last two decades, a significant volume of research was conducted targeting the coherent combination of arrays of lasers in order to gain understanding of laser dynamics, control speckle and achieve higher output power while maintaining a fixed phase relationship between multiple emitting laser cavities \cite{Cao2019,Khurgin:2005,Huang:2005,Huang:2009,Liu:2019}. Coherent arrays of lasers were extensively researched in the case of laser diodes as they are an ideal platform for coupled cavities \cite{Corcoran:05,Corcoran:14}. Equally, In-plane lasing in polariton lasers has been demonstrated, in which the gain is sufficient to interact with metallic gratings and electrode structures, while still supporting a polariton state \cite{Suarez-Forero:20}. The excitonic dipole in a semiconductor quantum well (QW) sample lies on the plane perpendicular to the growth direction of the QWs and parallel to the surface of the wafer, and therefore photonic emission and lasing is possible in the direction parallel to the surface plane. In-plane lasing has also been investigated previously as a very persistent parasitic effect in optically pumped vertical external-cavity surface emitting lasers (VECSELs) \cite{Hessenius:2011,Tino:12,Wang:15,Guina:2017}, which occurs due to the high gain from semiconductor QWs (on the order of 30\,dB/cm per QW) \cite{Corzine:90,Summers:95}. Here we use the same high-gain emission process to propose a laser modality that exhibits highly efficient coherent laser arrays and can be useful for future integration with other integrated photonic technologies.

In this work we present optically pumped membrane quantum well lasers (MQWLs), lasing in-plane without the use of an external cavity, and with sub-millimeter cavity lengths. The membrane gain medium is grown with metallo-organic vapour phase epitaxy (MOVPE) and transferred onto a silicon carbide (SiC) or oxidised silicon. The membrane geometry is used as it enhances the laser performance on account of the strong waveguide confinement created by the high refractive index of the membrane in comparison to the underlying substrate. Also, the underlying substrate serves as an efficient heat spreader because of the micron-thick membrane geometry. For small cavities, on the order of 70\,\textmu m from facet to facet we record lasing thresholds of approximately 60\,mW of pump power. We go on to demonstrate an array of lasing cavities which all operate coherently on a single common longitudinal mode in longer cavities (hundreds of  microns) on silicon carbide and oxidised silicon substrates. The coherence across the lasing cavities is further shown to be maintained even when the individual cavities operate simultaneously on multiple longitudinal modes. During in-plane lasing, where the optical pump is imaged on the surface of the membrane laser, the laser cavity positions and the coherence of the laser array can be controlled by the position and spatial intensity distribution of the pump laser.

MQWLs exhibit a number of advantages, some of which are similar to other epitaxially grown semiconductor membranes \cite{Zang:16,Kahle:16,Mirkhanov:17,Priante:21}: (i) the thin growth with released stresses exhibits excellent crystallinity, and hence reduced non-saturable losses; (ii) MQWLs can be grown in about a tenth of the time that is needed to grow a VECSEL or polariton laser, and therefore are an ideal platform to investigate new functionalities; (iii) MQWLs are an ideal geometry in order to extract heat by contact bonding to sapphire or SiC, and can therefore be highly efficient lasers; (iv) MQWL growth is freed from the material (lattice matching) constraints of DBR mirror growth, which can adversely affect the quality of growth (a typically observed characteristic of VECSELs and some polariton lasers); and therefore (v) MQWLs can be designed in a wider selection of design wavelengths as the lack of DBR liberates the development of lasers at wavelengths where appropriate DBR material index contrast is not possible, such as the all-important 1.5\,\textmu m telecoms wavelength region; and finally, (vi) because of the high index contrast between the membrane material system and the substrate (silicon carbide, silica etc.), strong waveguiding is formed which ensures a high overlap of the guided mode with the QW gain region.

Operating MQWLs in the in-plane waveguide geometry provides some unique possibilities for the future. Specifically, they are optically re-configurable via manipulation of the shape and position of the pump laser beam, and we demonstrate in this work that they can operate as a single laser or in coherent arrays. In future work, we can optimise the active region design of the MQWL presented here to increase the available gain by incorporating more QWs in the centre of the active region to increase overlap with the waveguide mode. Similarly we can decrease the waveguide losses due to scattering by using a double hetero-structure geometry where the active region is grown in a sandwich of lower refractive index semiconductors. While the pump laser in this work is a multimode fibre coupled diode laser, it is hoped in future that accurate control of the behaviour of the MQWL may be achieved using a single spatial mode pump in combination with a spatial light modulator. A knock on effect of the reconfigurable nature of MQWLs in our geometry is the ease by which they may be integrated with photonic structures:  MQWLs can be integrated as pump laser with other photonic technologies, such as: nonlinear waveguide systems, Kerr micro-combs \cite{Gaeta2019}, silicon photonics circuits \cite{Liu2018} and optical computing concepts such as neural networks based on multimode multiplexer interferometers \cite{Qian2020}.

\section{Sample description, Imaging system and calibration}

\begin{figure}[ht!]
    \centering
    \includegraphics[scale=0.55]{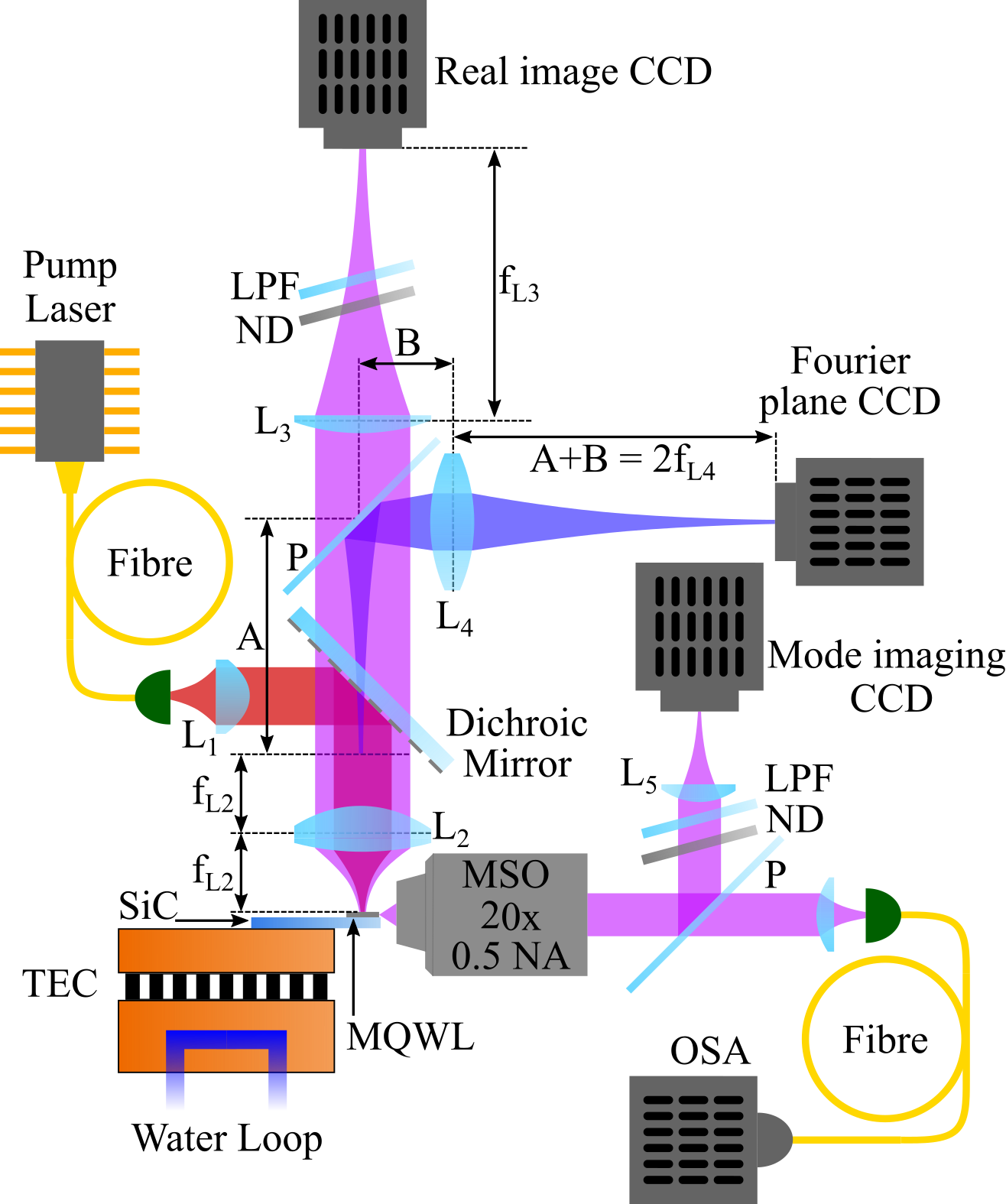}
    \caption{Schematic diagram of the basic membrane laser pumping and imaging experimental setup. The pump light is coupled to a 100\,\textmu m core diameter multimode fibre, and following emission to free space is collimated (f\textsubscript{L1}\,=\,25\,mm) and focused (f\textsubscript{L2}\,=\,30\,mm) onto the membrane sample. Real and reciprocal space imaging beams are obtained, respectively, by the transmission and reflection at a pelicle beam splitter. The real and reciprocal space beams pass through lenses L\textsubscript{3} (f\textsubscript{L3}\,=\,150\,mm) and L\textsubscript{4} (f\textsubscript{L4}\,=\,75\,mm) before detection on CCD cameras. A 20\texttimes~microscope objective is mounted such that the optical axis of the objective is coaxial with the waveguide axis of the laser cavity to allow for imaging of the waveguide mode intensity profile. The beam from the objective is split with a pelicle such that the reflected beam is imaged onto a CCD using L\textsubscript{5} (f\textsubscript{L5}\,=\,250\,mm), and the transmitted beam is coupled to a 50\,\textmu m diameter multimode fibre for measurement on an optical spectrum analyser (OSA). LPF\,=\,Long Pass Filter, ND\,=\,Neutral Density filter.}
    \label{fig:setup}
\end{figure}

The sample of semiconductor QW membrane used throughout this work is similar to a membrane external-cavity surface emitting laser (MECSEL) sample which is also a geometry of a membrane optically pumped QW laser. The thickness of our membrane is designed to be 1570 nm thick and consists of 10 InGaAs QWs. Each QW is comprised of 10\,nm of In\textsubscript{0.13}GaAs sandwiched between two 21\,nm barrier layers of GaAsP\textsubscript{0.06}. Repetitions of these three layers are then separated by 90.6\,nm of GaAs buffer layers. The two outermost GaAs buffer layers are 100\,nm thick, and two 20\,nm capping layers of GaIn\textsubscript{0.06}P then complete the membrane sample. The membrane is grown on a GaAs substrate, and between the substrate and the first capping layer is 200\,nm of AlAs which acts as an etch stop. The sample is the same geometry as the one used in \cite{Mirkhanov:17}. In preparing the membrane laser sample, the QW side of a sample chip is adhered to a piece of silicon wafer. The GaAs substrate is removed during wet processing using a solution of ammonium hydroxide with hydrogen peroxide. The AlAs layer is removed with a solution of hydrofluoric acid. The resulting semiconductor membrane is then released from its host substrate inside a solution of acetone, the acetone is slowly swapped to isopropanol taking care to always have the membrane floating in the solution. Consequently the membrane is further broken into small pieces, and then capillary bonded to the desired substrate simply by removing the substrate with the membrane on top out of the solution and letting it dry. In our case the host substrate is a silicon carbide or oxidised silicon (Boron doped) substrate, the fabrication steps are the same regardless of the substrate. The fabrication steps follow the same recipe as in MECSELs used in \cite{Kahle:16,Mirkhanov:17} and the contact bond is good enough to withstand 10s of Watts of pump power when silicon carbide or sapphire is used as a supporting substrate \cite{Mirkhanov:17,Kahle2019}.

The experimental setup for characterising the QW membrane lasers is given schematically in figure \ref{fig:setup}. The completed membrane sample is thermally contacted to a copper block whose temperature is maintained by a thermoelectric element and a PID servo at 14$^{\circ}$C. This value is chosen as it is the coolest temperature we can achieve in our lab before atmospheric water condensation begins to form. The thermoelectric element sinks heat transferred from the sample into a closed loop water cooled copper block. Where lasing in the sample occurs along the axis of the membrane that is perpendicular to the growth direction, it is most convenient for the pump light to be incident along the axis normal to the membrane surface. Realising this, while still maintaining the ability to image the top surface of the membrane (using the subsequent emission from the membrane) is achieved through the use of a dichroic mirror with a passband starting at 900\,nm. We are able to pump with up to 4\,W of 808\,nm pump light using a (multimode, 105\,\textmu m core diameter) fibre coupled diode bar, whereby light from the fibre is first collimated before being imaged onto the membrane surface. In the case where lenses L\textsubscript{1} and L\textsubscript{2} have focal lengths of 25\,mm and 30\,mm respectively, the resulting pump spot has a full width at half maximum of $\sim$100\,\textmu m.

Real space and reciprocal space imaging of the sample in the axis normal to the membrane surface is achieved using the pump focusing lens, L\textsubscript{2}, as the primary lens \cite{Maslov:04,Saxena:15}. Collected light from the membrane sample first passes the dichroic mirror before being split with a beam splitter in order to allow capture of real and reciprocal space images simultaneously. The pump spot is still visible in the resulting images, owing to the extinction ratio of the dichroic mirror. In the transmitted direction of the pelicle, a single plano-convex spherical lens, L\textsubscript{3} is in position to form the real space image on a CCD. In the reciprocal space imaging arm, L\textsubscript{4} is placed at the appropriate distance to image the reciprocal space of lens L\textsubscript{2} on to the CCD.

Light collection coaxial to the in-plane lasing cavity is achieved using a 20\texttimes, 0.5 NA microscope objective. The collimated beam is subsequently split into two using a 10\%(R)\,:\,90\%(T) pelicle. The weaker beam is further attenuated before being focused onto a CCD to provide imaging of the mode intensity profile at the membrane edge facet. The more powerful component is coupled to a multimode fibre and to an optical spectrum analyser (OSA). 

\begin{figure}[ht!]
    \centering
    \includegraphics[scale=0.45]{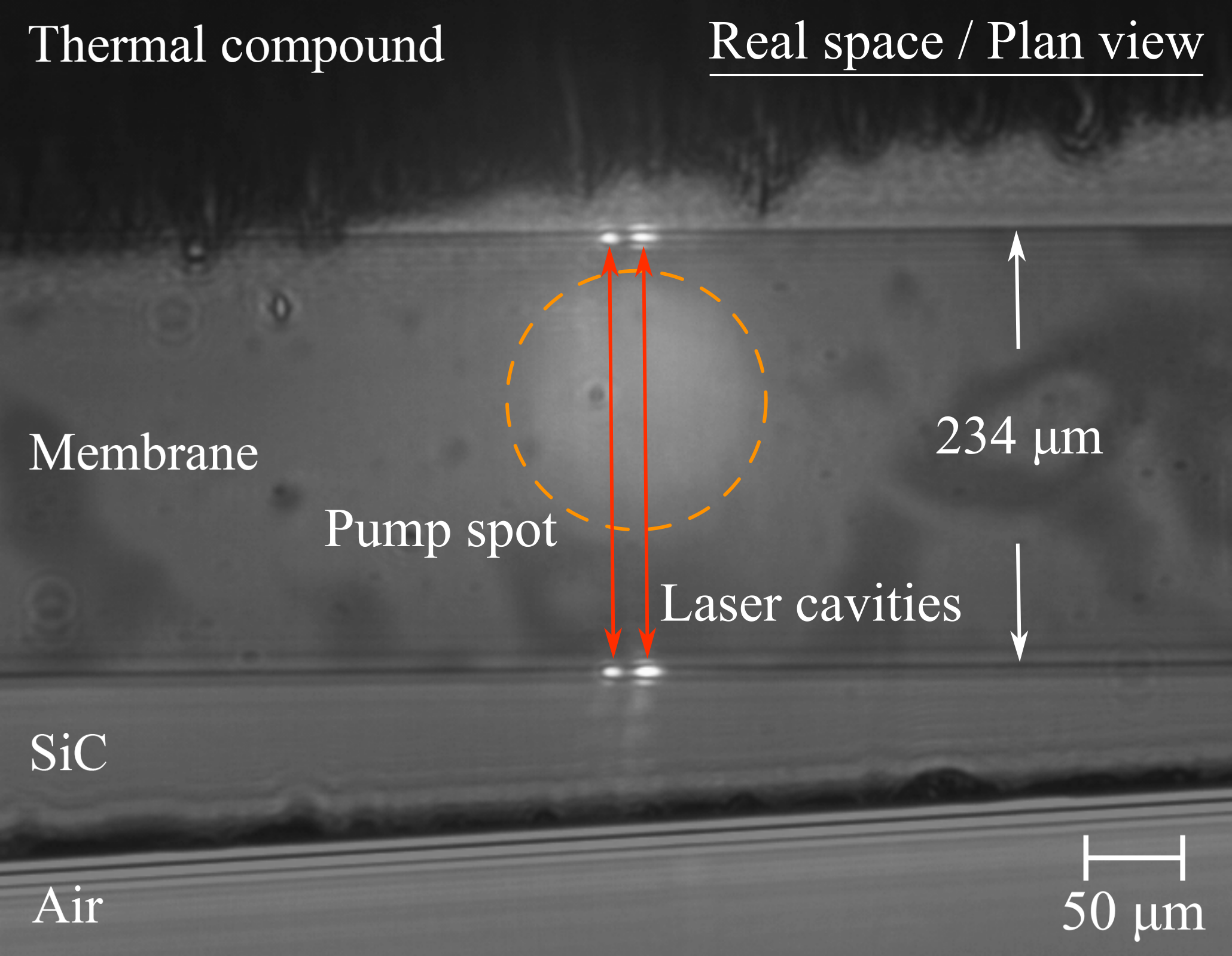}
    \caption{A hybrid (top view) image of the membrane taken on the real space imaging camera. The two images comprising this photograph consist of one image where the LPF was removed and where the pump running at approximately 10$\%$ of threshold current, and another image where the LPF is replaced and the pump is run above membrane threshold power. The two images are overlaid and the top layer opacity is set to 50$\%$. The black region at the top is the extent of the opaque heatsink compound between the copper block and the SiC heat spreader which has bled out from the interface (the membrane is bonded to the opposite face of the SiC), and the diagonal line at the bottom of the image is the edge of the SiC heat spreader. Lasing can be seen by the bright spots located at the edges of the membrane either side of the centre of the pump spot, and the lasing cavity length is 234\,\textmu m long.}
    \label{fig:membranehybrid}
\end{figure}

The imaging system must be calibrated against object (real space) and wavenumber (reciprocal space) of known size and periodicity respectively. Calibration is defined here as the conversion of pixels on the CCD to microns in real space, and pixels on the CCD to degrees in reciprocal space. In order to calibrate the imaging system, we use a metallic transmission electron microscope (TEM) grid with well defined physical dimensions. The TEM grid also acts as a transmission diffraction grating, and hence imaging the reciprocal space formed by the primary collection lens in the setup provides a method for calibrating the resulting diffraction patterns observed from the membrane samples. We use the reciprocal space image of the membrane laser as it reveals the laser threshold accurately: the instant that lasing starts, a diffraction pattern is observed. Furthermore, the membrane lasers operate into multiple laser cavities, as in the case depicted in figure \ref{fig:membranehybrid}, where two laser cavities can be seen operating closely together; the interference (or lack of interference) of the reciprocal space images is a strong indication for the coherence of the multiple adjacent laser cavities. We use degrees rather than wavenumbers for the reciprocal space, as we investigate the scattering from the end-facets as interference from multiple coherent point sources.

In figure \ref{fig:membranehybrid} we show a composite image of a silicon carbide mounted MQWL in operation.  The figure is comprised of two separate images of the membrane, one above lasing threshold and the other below, both captured using the real space imaging camera. In the latter image case, the sample is front illuminated with white light and the long pass filter is removed from the imaging setup. The two images are overlayed and the top layer opacity is dropped to 50\%. Lasing operation is revealed by the bright spots located at the membrane facets, caused by scattering of the emitted light at the facets. Two separate cavities are visible, which are located near the central region of the pump spot which is visible as the bright circular area just up from the geometric center of the cavity facets. At this time, the cause of the larger, softer edged, darker areas within the sample remains unknown. We speculate that the cause could be down to localised incomplete bonding of the membrane to the SiC heat spreader, or more possibly incomplete etching of the AlAs etch-stop layer during wet processing. Most of the data presented in this paper were made using one or more pumped locations on the sample described in figure \ref{fig:membranehybrid} which is an MQWL mounted on a silicon carbide. However, we also have used a smaller cavity sample again mounted on silicon carbide which exhibited thresholds down to 60 mW and a silicon mounted sample which is shown below in figure \ref{fig:siliconmounted}.

Finally, it is worth stating explicitly that we use the group refractive index, $n_{g}$, when calculating the Free Spectral Range (FSR) of our MQWL waveguide geometry. The group refractive index may be evaluated at a wavelength of 1\,\textmu m. We model our structure as a waveguide in COMSOL which yields a value of effective refractive index of $n_{e}$\,=\,3.49, and $\nicefrac{dn_{e}}{d\lambda}$\,=\,$-$0.65\,\textmu m\textsuperscript{-1}. Consequently, we arrive at the value for the group refractive index, $n_{g}$\,=\,4.14\,$\pm$\,0.11.

\section{Results}

\begin{figure}[ht!]
    \centering
    \includegraphics[scale=0.3]{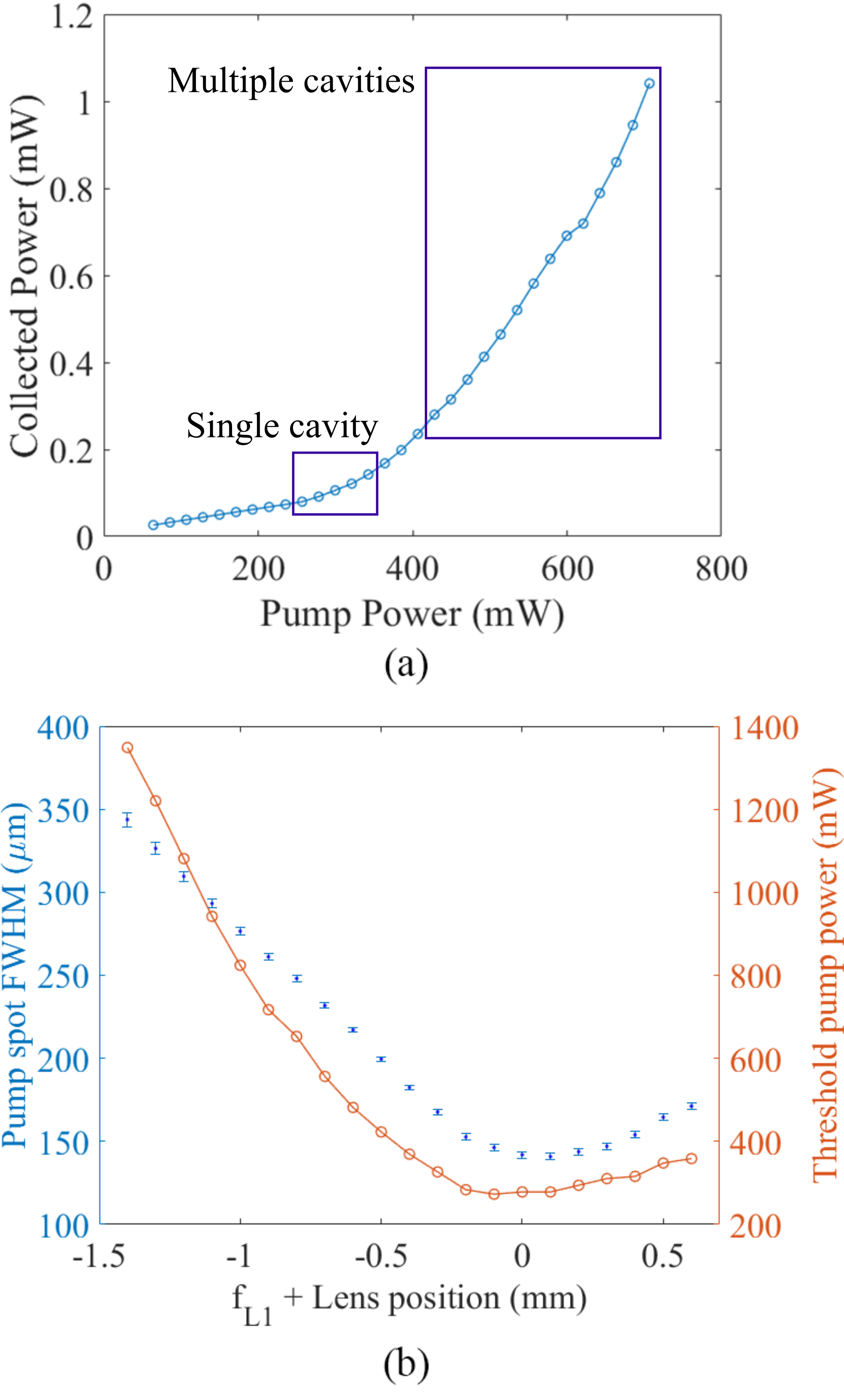}
    \caption{(a) Collected laser power as a function of incident pump power for a pump spot size of approximately 100\,\textmu m. (b) Comparison of the membrane laser threshold pump power as a function of the circular pump spot FWHM dimension. Breathing of the pump spot FWHM is achieved using a simple two lens imaging pair (pump collimation and pump focusing), whereby the collimation lens is translated around either side of the collimation condition when the lens is its own focal length away from the pump laser fibre tip.}
    \label{fig:lasercharac}
\end{figure}

Figure \ref{fig:lasercharac} demonstrates both threshold and efficiency behaviour, as well as the relationship of the pump spot FWHM and the membrane laser threshold for a specific pumped location on the silicon cabride mounted MQWL. The laser threshold, as can be seen in figure \ref{fig:lasercharac}a, is approximately 275\,mW when using a 25\,mm collimation and 30\,mm focusing lens. At the point of reaching threshold, only a single cavity located to approximately coincide with the center of the pump spot is in operation. Overlap of the optical mode with the pump spot in this case is poor as the pump spot FWHM is approximately an order of magnitude wider than the cavity mode, and hence the increase in gradient above threshold is marginal. With a small increase in pump power however, to approximately 425\,mW, multiple lasing cavities have reached threshold, and hence the efficiency of the system as a whole improves by nearly a factor of eight (depicted by the change of gradient in figure \ref{fig:lasercharac}a. In this study, we have not optimised the overlap of the pump spot with any given optical cavity mode within the membrane. It is anticipated that improving the overlap of the pump spot with the lasing mode will decrease the contribution of fluorescence to the collected power, reduce the membrane lasing threshold, and improve the external efficiency.

Figure \ref{fig:lasercharac}b demonstrates the measured trend of lasing threshold pump power with respect to pump spot FWHM. Here the collimation lens L\textsubscript{1} is translated along the beam axis to defocus the pump spot on the membrane. It is important that, in this two lens pump light imaging system, the first lens is moved rather than the second, and this is because the second lens position is associated with the imaging of the sample. Fitting to the fluorescence in the real space images provides the pump spot dimensions. We find that at the translation position where the lens is imaging the fibre facet (0 mm), the beam profile approximates a top hat beam profile as expected for a heavily multimode optical fibre. We see that the position where the lowest threshold power was recorded is at -0.2\,mm, which is slightly offset to the minimum FWHM position, this is because at the -0.2\,mm position we have slightly higher central pump peak intensity.

\begin{figure}[ht!]
    \centering
    \includegraphics[scale=0.18]{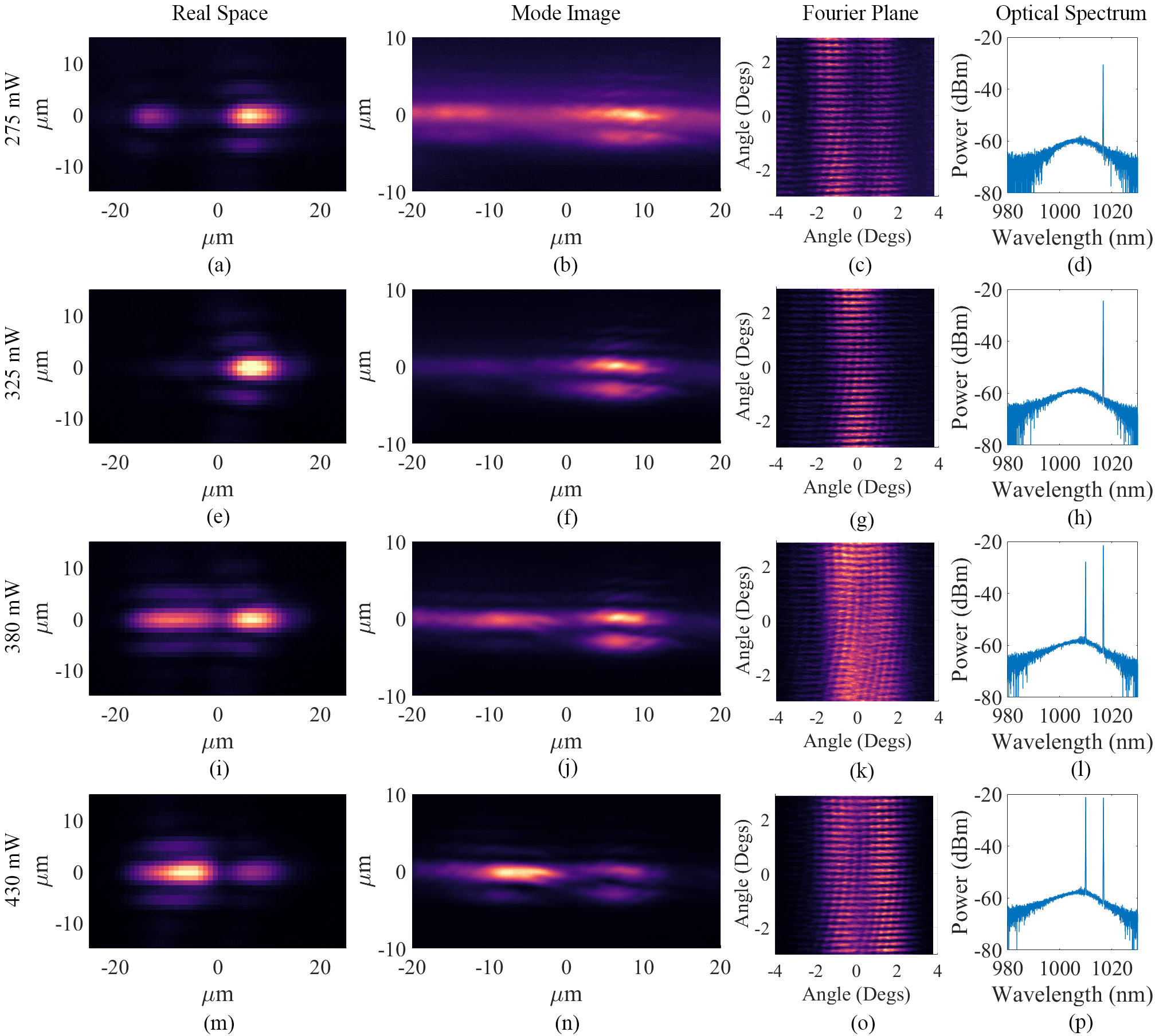}
    \caption{Images showing corresponding laser measurements at increasing pump powers. Each row (a - d, e - h, etc) represents the pump powers of 275\,mW, 325\,mW, 380\,mW, and 435\,mW respectively, and with the pump conditions stated above, the lasing threshold at this position on the sample is approximately 250\,mW. The first column represents the real space image (plan view) of one end facet of the MQWL measured from top down obtained with lens L\textsubscript{2} (a, e, i, m), column 2 shows an end-on view of the waveguide mode intensity of the MQWL (b, f, j, n), column 3 shows the imaging of the reciprocal space of the MQWL using lens L\textsubscript{2} (c, g, k, o), and finally colums 4 shows the measured optical spectra (d, h, l, p) respectively. All images are captured with monochrome cameras and the raw data is normalised to unity and mapped to a linear colour scale. The OSA resolution is set to 0.05\,nm. Specifically in the case of figure \ref{fig:modeimagesfourierosa}l and \ref{fig:modeimagesfourierosa}p, the wavelength separation in both subfigures is 6.8\,nm, or equivalently $\sim$2\,THz.}
    \label{fig:modeimagesfourierosa}
\end{figure}

In figure \ref{fig:modeimagesfourierosa} we compare the imaging of the real and reciprocal space of the same MQWL as viewed from the top, against the corresponding waveguide mode intensity images and optical spectra for increasing pump powers. Here, the importance of the reciprocal space imaging becomes clear as it enables a better understanding of the lasing conditions within the membrane. In the simplest case, the second row (figures\,\ref{fig:modeimagesfourierosa}e\,-\,h) shows one laser cavity with a single waveguide spatial mode as captured by the microscope objective, and a single peak in the optical spectrum. Correspondingly, the reciprocal space imaging shows an interference pattern in the vertical axis caused by the interference of the light from the facets at both ends of the cavity, and since it is only one laser cavity no signs of interference in the intensity of the horizontal axis, the diffraction pattern is centered around 0 degrees as expected. By contrast, the waveguide mode image and real space image in the first row (figures\,\ref{fig:modeimagesfourierosa}a\,-\,d) clearly show two peaks in the spatial intensity indicating two laser cavities in the MQWL, and yet only a single peak in the optical spectrum. Correspondingly the reciprocal space imaging now shows an interference pattern in the horizontal axis, confirming the spatially discrete light sources and therefore suggesting that the two axial cavities in the membrane coherently seed one another. An important observation here is the intensity peaks are positioned either side of zero degrees, indicative of the light fields in each discrete cavity being in anti-phase with one another. Indeed, as the resolution of the optical spectrum analyser is enough to resolve two adjacent axial modes of the same optical cavity, a reasonable conclusion is that the two optical cavities are operating coherently.

Fitting to the reciprocal space images in the vertical direction, in all four cases, yields a laser cavity facet separation of 251.9\,$\pm$\,3.0\,\textmu m, which corresponds to a FSR frequency of 144.0\,$\pm$\,5.7\,GHz if we use the group refractive index of 4.14\,$\pm$\,0.11. In the real space images the measured facet separation of the membrane laser is 234\,$\pm$\,5\,\textmu m, which corresponds to 154.7\,$\pm$7.2\,GHz (where the cavity length uncertainty is assumed to 2.5\,\textmu m per facet).

\begin{figure}[ht!]
    \centering
    \includegraphics[scale=0.2]{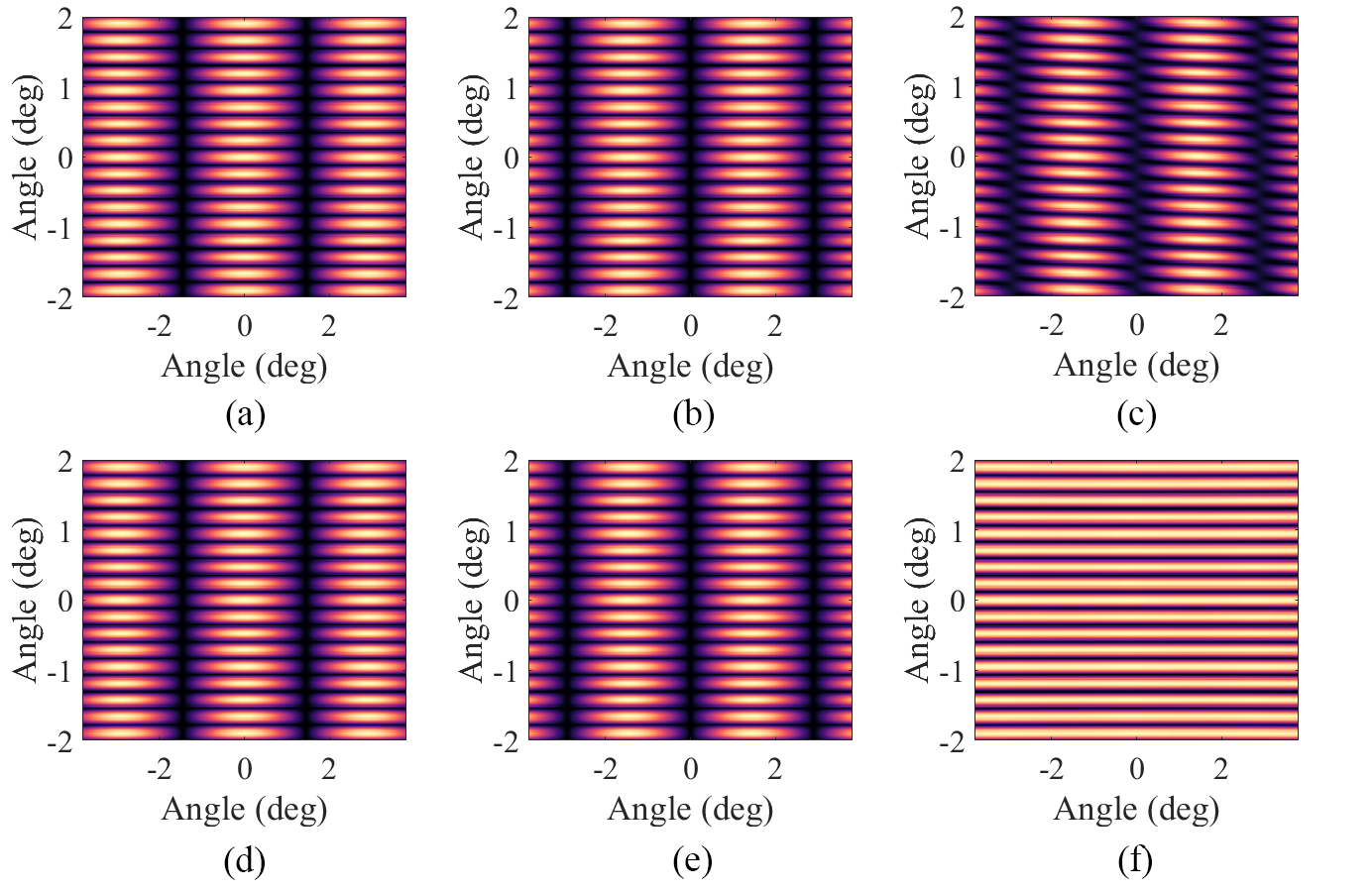}
    \caption{Numerical modelling of the far-field intensity of the interference pattern originating from two membrane axial cavities. The cavity length is 234\,\textmu m (measured from the real space images) and optical axis separation between the two is 20\,\textmu m. We arbitrarily denote one cavity as c\textsubscript{1} and the other as c\textsubscript{2}, the frequency which each operates on as \textomega\textsubscript{c\textsubscript{1}} and \textomega\textsubscript{c\textsubscript{2}}, and the relative phase difference between each cavity as \textphi. In the subfigures, we model the cavity configurations as (a) \textomega\textsubscript{c\textsubscript{1,2}}\,=\,1016.7\,nm, \textphi\,=\,0; (b) \textomega\textsubscript{c\textsubscript{1,2}}\,=\,1016.7\,nm, \textphi\,=\,\textpi; (c) \textomega\textsubscript{c\textsubscript{1,2}}\,=\,1016.7\,nm, \textphi\,=\,\textpi, and where a single emitter of the four has a factor of three increase in field amplitude; (d) \textomega\textsubscript{c\textsubscript{1,2}}\,=\,1010.0\,nm, \textomega\textsubscript{c\textsubscript{1,2}}\,=\,1016.7\,nm, (both cavities simultaneously operating on both optical frequencies) \textphi\,=\,0; (e) \textomega\textsubscript{c\textsubscript{1,2}}\,=\,1010.0\,nm, \textomega\textsubscript{c\textsubscript{1,2}}\,=\,1016.7\,nm, \textphi\,=\,\textpi; and (f) \textomega\textsubscript{c\textsubscript{1}}\,=\,1010.0\,nm, \textomega\textsubscript{c\textsubscript{2}}\,=\,1016.7\,nm. In the case of subfigures d\,-\,f, we perform the average over 1000 optical wavelength periods in order to indicate what would be experimentally observed on the timescale of the CCD exposure time which is far greater than the time for the relative phase between the two optical cavities to precess around 2\textpi.}
    \label{fig:modellingfigure}
\end{figure}

Figures\,\ref{fig:modeimagesfourierosa}i\,-\,l and figures\,\ref{fig:modeimagesfourierosa}m\,-\,p, exhibit two cavities operating in the real space images, a horizontal periodicity in the x\,axis of the reciprocal space images revealing interference between the adjacent cavities, and two distinct wavelength components in the optical spectrum. The reciprocal space images all show minima around 0 degrees in the x axis, supporting the notion that the longitudinal cavities are operating \textpi~out of phase with one another. It is proposed that collectively in these cases, all optical cavities present must be operating coherently on both spectral components present on the optical spectrum trace. We discuss this condition in further detail below in conjunction with figure \ref{fig:modellingfigure}.

Figure \ref{fig:modellingfigure} represents the numerically derived reciprocal space image (or far field interference pattern) for the experimental conditions depicted in figure \ref{fig:modeimagesfourierosa} for various optical conditions (see figure caption of \ref{fig:modellingfigure} for more details). The model is very simple and calculates and plots the 3D interference on a Cartesian plane of 4 point sources that represent both ends of the laser cavities of the MQWL operating side by side. The 4 point sources represent the 4 bright spots that show the scattering of the laser end facets shown in figure \ref{fig:membranehybrid}. In all cases we model two optical cavities which are 234\,\textmu m in length, and separated by 20\,\textmu m, based on our real space imaging. The first row (figs \ref{fig:modellingfigure}a\,-\,c) considers both cavities operating on identical optical frequencies, while in the second row (figs \ref{fig:modellingfigure}d\,-\,f) considers multiple frequency components in the optical spectrum. In the first instance, figures \ref{fig:modellingfigure}a and \ref{fig:modellingfigure}b show the salient difference for a mutual phase difference between two cavities operating on identical wavelengths of 0 and \textpi, respectively. A 0 phase difference gives a maximum on the origin of the horizontal axis whereas a \textpi~phase difference shows a minimum at the origin confirming that the cavities presented on figure \ref{fig:modeimagesfourierosa} work with \textpi~phase difference.  

When imaging the MQWLs from the top we image only the scattering at the end-faces of the laser cavities which is a random process and may not always represent the intracavity power of the laser. In order to explain reciprocal space images that show behaviours such as in figure \ref{fig:modeimagesfourierosa}(k), where there is a phase jump between the two interference patterns, we consider in the interference model the effect of altering the wave amplitude of a single facet of a single cavity in figure \ref{fig:modellingfigure}c. Here we set the amplitudes of the 4 point sources to be equal to A(c\textsubscript{1,top}, c\textsubscript{2,top}, c\textsubscript{2,bottom})\,=\,1, A(c\textsubscript{1,bottom})\,=\,3, which has the effect of rotating the lobes in the vertical periodicity clockwise and also introducing a phase discontinuity in the horizontal axis, very similar to what we see in figure \ref{fig:modeimagesfourierosa}(k). 

The second row of figure \ref{fig:modellingfigure} considers not only the presence of multiple frequency components, but also the time averaged nature of the imaging system used in the experiment. In the case of figures \ref{fig:modellingfigure}d and \ref{fig:modellingfigure}e, we model the presence of two optical wavelengths present in both cavities simultaneously (spaced by approximately 7\,nm to match the separation observed in the experiment and shown in figure \ref{fig:modeimagesfourierosa}), and set the phase difference between the two cavities as 0 and \textpi, respectively. We note a slight broadening of the vertical interference pattern caused by the presence of an additional, yet similar wavelength, but otherwise the behaviour is essentially identical to that of \ref{fig:modellingfigure}a and \ref{fig:modellingfigure}b respectively. Arguably the most important result from this modelling, is the scenario where one optical cavity operates on one optical wavelength while the neighbouring cavity operates on a different wavelength, as depicted in figure \ref{fig:modellingfigure}f. In this case, the relative phase between the cavities is constantly precessing at a rate equal to $1/(\omega_{H}-\omega_{L})$ cycles per second, where $\omega_{H}$ and $\omega_{L}$ represent the higher and lower optical frequencies respectively. From figures \ref{fig:modeimagesfourierosa}l and \ref{fig:modeimagesfourierosa}p, the absolute frequency difference is approximately 2.0\,THz, and hence it takes approximately 150 cycles of the fast optical frequency for the phase angles to realign. We therefore average over 1000 optical cycles to ensure an indicative result. In the hypothetical case where the two neighbouring cavities operate on different optical frequencies, it is evident that the horizontal periodicity of the interference pattern vanishes, and hence we conclude that it must be the case that, in figures \ref{fig:modeimagesfourierosa}l and \ref{fig:modeimagesfourierosa}p, both optical cavities coherently operate on all optical frequency components present in the optical spectrum.

\begin{figure}[ht!]
    \centering
    \includegraphics[scale=0.275]{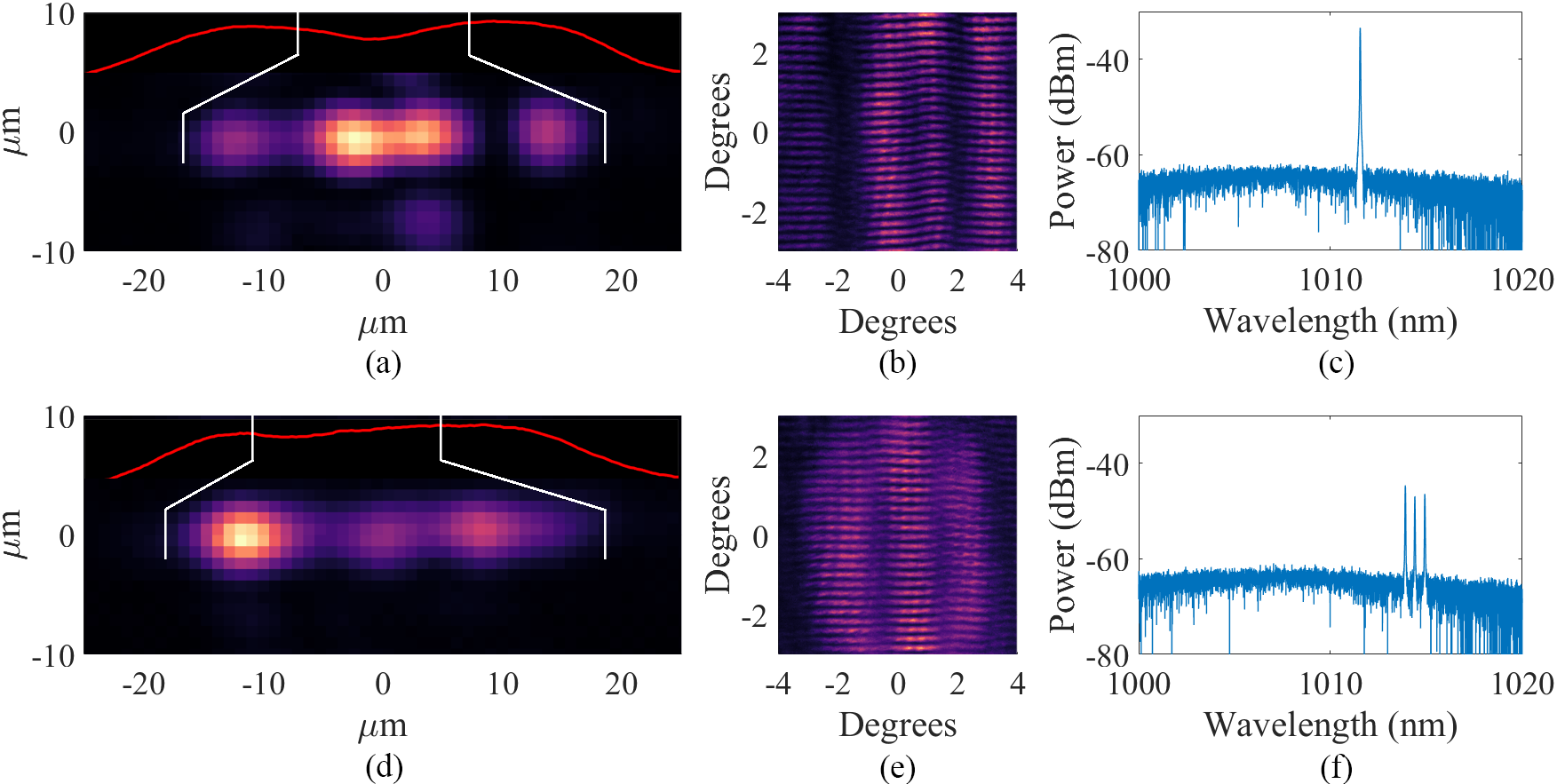}
    \caption{Figure showing the real space and reciprocal space images, as well as corresponding optical spectra, for two different lasing (pumped) positions on the membrane sample. The top row (a - c, $\sim$\,308\,mW pump) and bottom row (d - f, $\sim$\,327\,mW pump) show, respectively, that at two separate locations the membrane is capable of operating on single, and multiple optical wavelengths simultaneously, while in both cases operating on multiple lasing cavities. The red traces overlaid in subfigures (a) and (d) are single line data (smoothed) across the center of the circular pump spot fluorescence as viewed by the real space camera, and shown to the extent where the (band pass filtered) brightness falls to approximately the surrounding background level. The presence of the membrane lasing visibly depletes the carrier population within the pump spot, and hence shows that in (a) the membrane cavity is approximately centered in the pump spot, while in (d) the main lasing cavity is offset. In subfigure (f), the wavelength separation of each nearest neighbour pair is 0.496\,nm (144.6\,GHz).}
    \label{fig:modeimagesfourierosaarray}
\end{figure}

Figure \ref{fig:modeimagesfourierosaarray} demonstrates that MQWLs are capable of both operating on a common frequency across multiple lasing cavities, as well as on multiple frequencies per lasing cavity. In the case of the latter, due to the presence of the periodicity in the interference laterally in the reciprocal space images, all of the optical cavities are believed to be mutually coherent. The principle change between the top and bottom rows of subfigures is the pumped location on the membrane. 

Generally we find that along opposing edge facets which appear (under the real space imaging camera) to be well cleaved with no signs of defects, the majority of the membrane area will lase under sufficiently high pump power ($\geq$\,600\,mW). Specifically in the case of figure \ref{fig:modeimagesfourierosaarray} however, we aimed to demonstrate two modes of operation under very similar pump powers, in order to emphasize the re-configurable nature of the platform. In subfigures \ref{fig:modeimagesfourierosaarray}a - c, at the first position, it is clear from the real image that there are four lasing cavities within a 30 \,\textmu m span of the end facet. Correspondingly there is only a single peak in the optical spectrum, suggesting the cavities are all oscillating on this one frequency, figure \ref{fig:modeimagesfourierosaarray}b is showing three dominant structures in the horizontal axis, there is diminished fourth pattern in the left and this maybe the case as the scattering from this cavity maybe weaker. Conversely at the second position in subfigures \ref{fig:modeimagesfourierosaarray}d - f, three optical cavities are visible over approximately a 20 \,\textmu width, and three distinct peaks in the optical spectra. According to the modelling results shown above we believe that all cavities work coherently on all frequencies but of course the situation here is a bit more complex. In each of the real space images, the red trace overlaid shows the (central) profile of the pump spot PL as viewed by the real space camera, and the white lines depict where the lasing cavities sit relative to the pump spot. Owing to carrier depletion within the lasing volume, the dips in the fluorescence correlate with the alignment of the membrane lasing cavities within the pump spot.

In the case of figure \ref{fig:modeimagesfourierosaarray}f, the wavelength separation is 0.496\,nm for both cases of nearest neighbour pairs. Converted to a frequency, and using the central wavelength, this separation equates to 144.6\,$\pm$\,7.5\,GHz (where the uncertainty is assumed to approximately equal to the resolution limit of the spectrometer). The FSR of a linear cavity is calculated using the modelled group refractive index of 4.14\,$\pm$\,0.11 and the measured facet separation of 234\,$\pm$\,5\,\textmu m to be 154.7\,$\pm$7.2\,GHz (where the cavity length uncertainty is assumed to 2.5\,\textmu m per facet).
A different MQWL sample from the same wafer was investigated within a very similar setup to the one shown here. The lasing spectra show that this sample has a much lower threshold, and hence lasing can be observed with only 60\,mW of pump power. This is due to the shorter cavity length of approximately 70\,\textmu m. 

\begin{figure}[h!]
    \centering
    \includegraphics[scale=0.2]{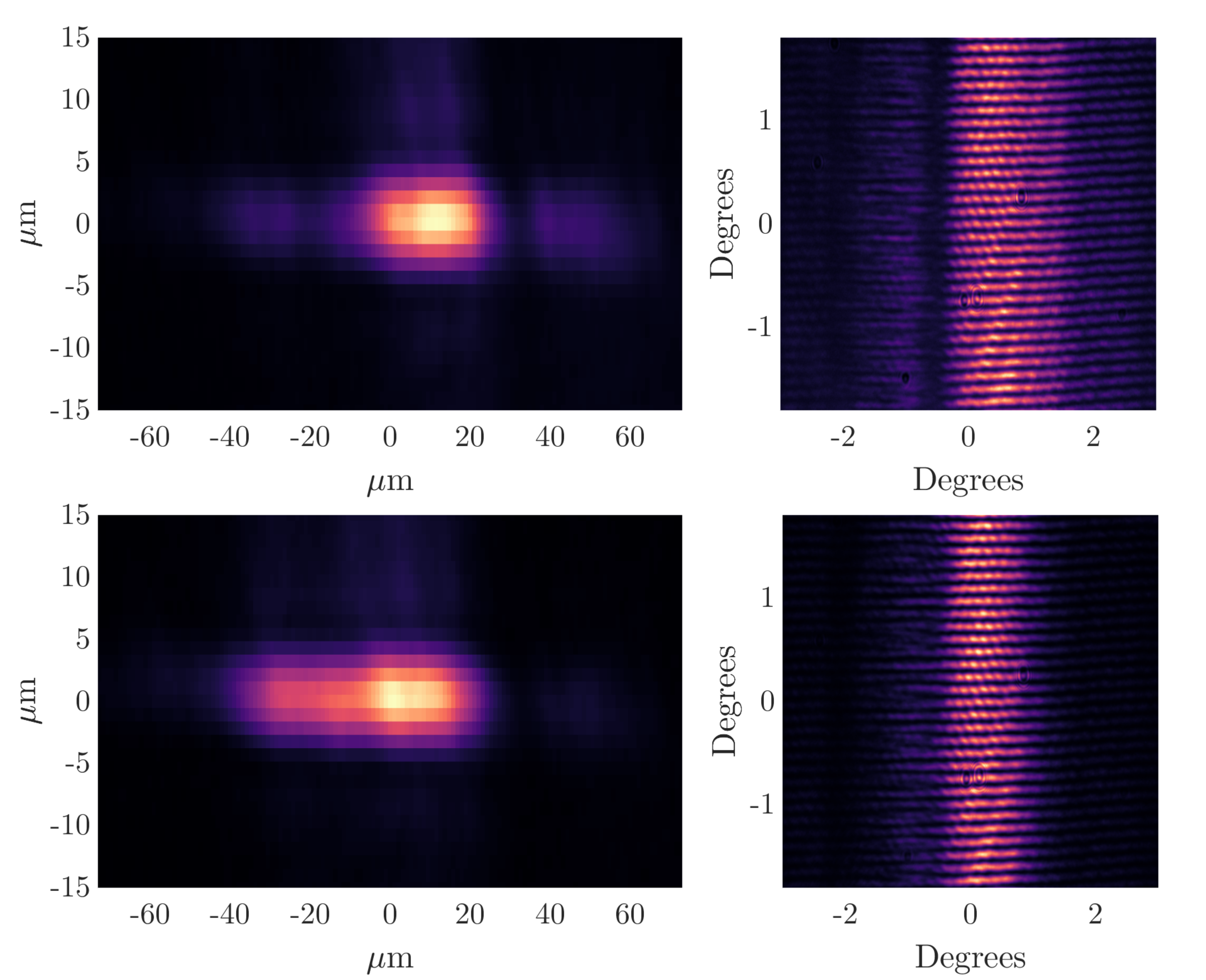}
    \caption{Figures showing the real space and reciprocal space images of an MQWL laser array on top of an oxidised silicon substrate. The real space images show only one end facet of the MQWL imaged from the top with lens L\textsubscript{2}. The transition from the first row to the second row happens by increasing the radius of the pump spot by translating the pump laser collimation lens (L\textsubscript{1}). The pump power is 251\,mW.}
    \label{fig:siliconmounted}
\end{figure}
In order to ascertain how appropriate the MQWL technology would be for applications with integrated technologies, we have tested an MQWL on a silicon substrate. In figure \ref{fig:siliconmounted} lasing can be seen in a 387\,\textmu m cavity mounted on a p-doped oxidized silicon substrate with 300\,nm of thermal oxide, the pump laser threshold is 211\,mW. In figure \ref{fig:siliconmounted} it can be seen that we can change the width of the waveguide lasing mode as we control the width of the laser pump spot using the collimation lens, L\textsubscript{1}. The reciprocal space images show the opposite behaviour as expected; the silicon mounted lasers always work at the same longitudinal mode at approximately 1017\,nm. The silicon mounted MQWLs can exhibit laser arrays for higher pump powers and they always operate at the same longitudinal mode, we have tested these lasers up to 600\,mW of pump power. The fact the MQWLs work in oxidised silicon is expected: the same Laser design, in an external cavity configuration, has been used in \cite{Mirkhanov:17} with a pump power of 40\,W and emitting up to 10\,W. At these low pump powers we believe that substrates such as oxidised silicon give enough heat dissipation for the operation of the MQWL. The performance of the MQWLs at low powers is more dependant on the quality of the bond of the membrane to the substrate rather than the heat conductivity of the substrate. If there is need to go to a considerably higher power then a substrate such as SiC or sapphire (which is also integration compatible) should give much higher thermal rollover point and higher output power. It is unlikely that this kind of performance would be needed for integrated photonic applications that we target here.


\section{Discussion}

From the presented experimental results and modelling, it can be seen that the interference shown in the x-axis of the reciprocal space images reveals that the lasing cavities are coherent with one another. Were this not the case, and the lasers were operating at different frequencies, then the far-field interference (or reciprocal space) imaging patterns would resemble figure \ref{fig:modellingfigure}(f). In figure \ref{fig:modellingfigure}(d) the far-field interference of two coherent lasers operating at the same frequency and relative phase is shown, and comparing this figure with our experimental results as shown for example in figure \ref{fig:modeimagesfourierosa}, we may conclude that our lasers are always operating at a \textpi~phase difference between them. In the case of figure \ref{fig:modellingfigure}(f), each cavity operating on mutually exclusive optical frequencies eliminates the horizontal periodicity observed in both figure \ref{fig:modeimagesfourierosa}(c) and \ref{fig:modeimagesfourierosa}(o). The experimental results shown in figure \ref{fig:modeimagesfourierosa} therefore demonstrate that it must be the case that both optical cavities are simultaneously operating on both wavelength components present in the optical spectrum, and that a \textpi~phase shift exists between them. The lasers continue to be coherent even when operating in multiple longitudinal modes.

Regarding the position of the lasing cavities (for example in figure \ref{fig:modeimagesfourierosaarray}), it is observed that, while within a region of the pump spot, not only are the positions of the lasing cavities within the membrane itself repeatable between pump laser power cycles, but also that if they are well above threshold intensity, the cavities will translate with the membrane if the global position of the membrane is changed using the kinematic translation stages. It was found that this behaviour was repeatable despite varying the temperature and the exact pump position. Consequently, it is concluded that the initial in-plane lasing cavity positions are governed most fundamentally by the positions in the membrane which are subject to the least loss and therefore a lower threshold. Spatial variations of loss in the samples can be attributed to the quality of the edge facets, scattering loss from imperfections on the surface and from e.g. growth defects or regions where the MQWL is not adequately bonded to the SiC heatspreader. The width of the laser waveguide modes in all our experiments are in the region of 5\,-\,10\,\textmu m wide, whereas the pump spot has a radius of 100 \textmu m. This is because the circular shape of the pump laser favors lasing at the middle region of the pump. Also, a narrow laser region can have lower losses and therefore a lower laser threshold than a laser with a wide guiding mode that would average out many defects on the membrane and end-facets. A narrow laser region may be also favorable because it draws carriers from the surrounding pumped area keeping a smaller ratio of the area that carriers are depleted to the area that has available carriers. As we increase pump power, a second or a third laser in the array starts lasing and it is more efficient to be coherent at a \textpi~phase difference rather than at 0 phase difference. This is because the standing wave of waveguide laser cavities at \textpi~phase with one another permit for more efficient extraction of available carriers, laser cavities are "slotting" spatially creating standing waves at half wavelength phase difference between them. 

Assuming a loss on the order of 10\,dB/cm, and a reflectivity given from the Fresnel reflectivities of the end facets, we estimate the finesse of the cavity to be 2. From here, the Lorentzian resonance width is approximately 80\,GHz, which is half the FSR of the cavity of 154.7\,GHz. The finesse of our cavity and the performance of the laser can be improved, in future by: (i) lowering the propagation losses by making a heterostructure where the waveguide mode will be better confined and not overlapping with the surface interfaces, (ii) increase the end-facet reflectivity by coating the end-facets with reflective coatings or structuring the membrane with a grating to create a DFB laser geometry which could also lead to wavelength selectivity. Increasing the finesse of the laser will give us a narrower linewidth, as these lasers can work in single longitudinal mode, in future this can be an ideal laser platform for providing frequency stabilised laser operation.

\section{Conclusions}
We demonstrate waveguide membrane lasing in small cavities and correspondingly low pump power thresholds, most importantly we demonstrate lasing on a silicon substrate. These characteristics show a flexible high gain laser platform that can be readily integrated to other photonic technologies. We showcase here the potential of the platform and we believe that newly designed lasers can exhibit lower thresholds, higher power outputs, higher slope efficiency and narrow linewidths by: (a) re-positioning the QWs  to overlap more strongly with the higher mode intensity near the center of the waveguide, (b) shaping the pump to better overlap with the laser cavity, and (c) opt for double heterostructure designs which would lead to significantly reduced losses on account of improved mode confinement within the active region. In future work, the position of the pump laser will be controlled using a spatial light modulator that will enable control over the number of laser cavities and their relative coherence. Future work will also investigate creation of membrane samples of defined shapes and dimensions by way of microstructuring in order to investigate lasing geometries such as whispering galleries. We propose that this research could allow the development of arrays of lasers with tunable laser geometries, controllable coherence and could make an ideal source to interface with silicon photonics to be used for applications such as optical computing. 

\section*{Funding}
We would like to thank EPSRC for the grant, EP/T001046/1 titled "UK National Quantum Technology Hub in Sensing and Timing".

\section*{Acknowledgments}

\section*{Disclosures}

The authors declare no conflicts of interest.

\bibliographystyle{unsrt}
\bibliography{bib}

\end{document}